# TRANSFORMATION OF OPTICAL VORTEX BEAMS BY HOLOGRAMS WITH EMBEDDED PHASE SINGULARITY

A.Ya. Bekshaev[*], O.V. Orlinska

*I.I. Mechnikov National University, Dvorianska 2, 65082, Odessa, Ukraine*

**Abstract**

Spatial characteristics of diffracted beams produced by the "fork" holograms from incident circular Lauerre-Gaussian modes are studied theoretically. The complex amplitude distribution of a diffracted beam is described by models of the Kummer beam or of the hypergeometric-Gaussian beam. Physically, in most cases its structure is formed under the influence of the divergent spherical wave originating from the discontinuity in the beam spatial profile or its derivatives caused by the hologram's groove bifurcation. Presence of this wave is manifested by the ripple structure in the near-field diffracted beam profiles and in the power-law amplitude decay at the beam periphery. Conditions when the divergent wave is not excited are discussed.

The diffracted beam carries a screw wavefront dislocation (optical vortex) whose order equals to algebraic sum of the incident beam azimuthal index and the topological charge of the singularity imparted by the hologram. The input beam singularity can be healed when the above sum is zero. In such cases the diffracted beam can provide better energy concentration in the central intensity peak than the Gaussian beam whose initial distribution coincides with the Gaussian envelope of the incident beam. Applications are possible for generation of optical-vortex beams with prescribed properties and for analyzing the optical-vortex beams in problems of information processing.

*PACS*: 42.25.Bs; 42.25.Fx; 42.40.Jv; 42.50.Tx; 42.60.Jf; 42.90.+m

*Keywords*: Optical vortex; Computer-synthesized hologram; Transformation; Spatial structure; Beam propagation

Computer-generated holograms (CGH) with the groove bifurcation ("fork") are widely used for generation of beams with optical vortex (OV) [1–5] from an incident paraxial beam with a regular wavefront structure [6–10]. If a single groove divides into $\mu + 1$ branches (in Fig. 1 $\mu = 1$), the $n$-order diffracted beam acquires the OV with topological charge

$$l = \mu n. \qquad (1)$$

Integer number $\mu$ is usually referred to as topological charge of the phase singularity "embedded" in the CGH [11–13]; both $\mu$ and $n$ can be positive or negative. Properties of the OV beams produced by such a CGH were studied in detail for the most popular case when the incident beam is Gaussian [9–13]. However, the CGH of this type can also be applied for transformation of beams with already existing OV; then quantity (1) is algebraically added to the initial topological charge of the incident beam so the CGH acts as an OV transformer rather than an OV generator. In particular, such transformation can be employed for analyzing the OV beams, for example, those appearing in experiments with photons in the entangled states of orbital angular momentum [14,15] and for encoding and analyzing the information with assistance of OV beams [16]. Elaboration and further development of corresponding approaches require knowledge of special features and properties of the involved beams, and their investigation is the task of this paper.

---

[*] *Corresponding author. Tel.: +38 048 723 80 75
E-mail address*: bekshaev@onu.edu.ua (A.Ya. Bekshaev)

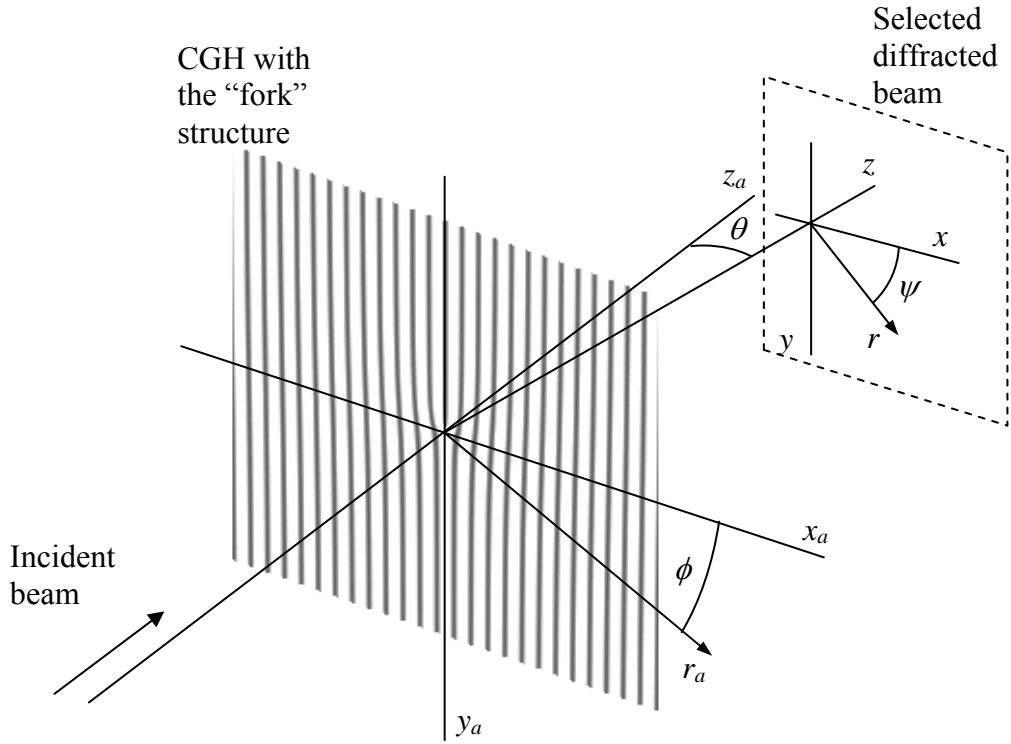

Fig. 1. Geometrical conditions of the beam transformation in the CGH (explanations in the text).

As is known, the spatial structure of the diffracted beams produced by the "fork" CGH essentially depend on many conditions, determining the diffraction regime: relative disposition of the CGH and the incident beam, diffraction order, angle of diffraction, etc. [12,13]. In this paper we restrict ourselves by the ideal "perfectly aligned" situations when the incident beam axis is orthogonal to the grating plane and intersects it exactly in the center of the hologram pattern (the groove bifurcation point). Besides, in agreement with the usual practice we assume that actual diffraction angles $\theta$ are small enough so that $\cos\theta \approx 1$ (see Fig. 1). This requirement implies that the considered CGH has low spatial frequency (practically, below 100 grooves per millimeter), and only low diffraction orders are admissible.

Our consideration is based on the general scheme of the beam transformation in the "fork" CGH developed in Ref. [12]. Geometrical conditions of the transformation are presented in Fig. 1. The CGH is considered as a planar transparency with spatially inhomogeneous transmittance $T(\mathbf{r}_a)$ where $\mathbf{r}_a = (x_a, y_a) = (r_a\cos\phi, r_a\sin\phi)$ is the radius-vector; the coordinate frame is chosen so that its origin coincides with the bifurcation point and axis $y_a$ is parallel to the grating grooves far from the "fork" (see Fig. 1). The transmittance function can be represented by the Fourier series,

$$T(\mathbf{r}) = \sum_{n=-\infty}^{\infty} T_n \exp\left[in\left(\frac{2\pi}{d}x_a + \mu\phi\right)\right]. \qquad (2)$$

where $d$ is the grating period. The incident beam is monochromatic with the wave number $k$; its axis coincides with axis $z_a$ that forms a 3D Cartesian frame together with axes $x_a$ and $y_a$ (at the grating plane $z_a = 0$).

Behind the grating, in accordance with expansion (2), a set of diffracted light beams (diffraction orders) is formed, propagating in directions specified by condition [17]

$$\sin\theta = \frac{2\pi n}{kd}. \qquad (3)$$

In fact, $n$-th term of expansion (2) produces the $n$-th diffraction order. To describe the field of a separate diffraction order, it is suitable to introduce the associated coordinate frame ($x, y, z$) with the origin in the bifurcation point (in Fig. 1, axes $x$ and $y$ are translated along axis $z$ to the current cross section). Corresponding polar coordinates are determined as usual,

$$r = \sqrt{x^2 + y^2}, \quad \psi = \arctan\left(\frac{y}{x}\right).$$

As was shown before [11,12], a diffracted beam of a separate order propagates along its axis $z$ as a paraxial beam and its field can therefore be represented as $E(x,y,z) = u(x,y,z)\exp(ikz)$ with the slowly varying complex amplitude $u(x,y,z)$ [18]. Under accepted small-angle diffraction conditions, the complex amplitude of the $n$-th diffraction order is determined by equation [12]

$$u_l(r,\psi,z) = \frac{k}{2\pi i z}\int u_a(r_a,\phi)\exp(il\phi)\exp\left\{\frac{ik}{2z}\left[r_a^2 + r^2 - 2r_a r\cos(\phi-\psi)\right]\right\} r_a\, dr_a\, d\phi \qquad (4)$$

where $l$ is defined by Eq. (1) and $u_a(r_a, \phi)$ is the complex amplitude distribution of the incident beam at $z_a = 0$. Of course, Eq. (4) determines $u_l(r,\psi,z)$ only within a constant factor; its absolute value should be found by another procedure taking into account the CGH diffraction efficiency (for example, corresponding coefficient in the Fourier expansion of the CGH transmission function (2)). For our present purposes, this efficiency is not important and for the determinacy, in Eq. (4) it is supposed to equal the unity, as if the whole input beam energy is transmitted to the chosen output beam.

Before proceeding further, note that for any paraxial beam, the two characteristic scales exist [5]. Let the characteristic scale of the beam transverse inhomogeneity be $b$; then the corresponding longitudinal scale is

$$z_R = kb^2.$$

With employment of these natural units of the beam geometric gauge, Eq. (4) takes on the simplified dimensionless form

$$u_l(\rho,\psi,\zeta) = \frac{1}{2\pi i\zeta}\int u_a(\rho_a,\phi)\exp(il\phi)\exp\left\{\frac{i}{2\zeta}\left[\rho_a^2 + \rho^2 - 2\rho_a\rho\cos(\phi-\psi)\right]\right\}\rho_a\, d\rho_a\, d\phi \qquad (5)$$

where $\rho_a = r_a/b$, $\rho = r/b$, $\zeta = z/z_R$.

In usual schemes where the CGH is used for imparting the screw wavefront to a regular beam, the incident beam complex amplitude distribution $u_a(\rho_a,\phi)$ possesses no phase singularity (for example, it was taken in the form of a Gaussian beam [12]). Now, keeping in mind our aim to study the CGH-induced transformations of beams with already existing OV, we take $u_a$ in the form of the standard OV model – a Laguerre-Gaussian mode [1,3] $LG_{pm}$ with zero radial index ($p = 0$) and arbitrary azimuthal index $m$ specifying the topological charge of the incident OV,

$$u_a(\rho_a,\phi) \equiv u_{am}(\rho_a,\phi) = \frac{\rho_a^{|m|}}{\sqrt{|m|!}}\exp\left(-\frac{\rho_a^2}{2} + im\phi\right). \qquad (6)$$

For this beam, $b$ coincides with radius of the Gaussian envelope measured at the level $e^{-1}$ of maximum and $z_R$ is the corresponding Rayleigh range [3]. The normalization constant $(|m|!)^{-1/2}$ in (6) warrants that all the considered beams carry the same total power regardless of $m$.

Substitution of (6) into Eq. (5) gives

$$u_l(\rho,\psi,z) \equiv u_{lm}(\rho,\psi,z) = \frac{1}{2\pi i\zeta}\frac{1}{\sqrt{|m|!}}\exp\left(\frac{i}{2\zeta}\rho^2\right)$$

$$\times \int_0^\infty \rho_a^{|m|+1} \exp\left[-\frac{\rho_a^2}{2}\left(1-\frac{i}{\zeta}\right)\right] d\rho_a \int_{-\pi}^{\pi} \exp\left[i(l+m)\phi - i\frac{\rho\rho_a}{\zeta}\cos(\phi-\psi)\right] d\phi \qquad (7)$$

In fact, this equation describes the free propagation of a paraxial beam with initial complex amplitude distribution

$$u_{lm}(\rho,\psi,0) = \frac{1}{\sqrt{|m|!}} \rho^{|m|} \exp\left[-\frac{\rho^2}{2} + i(m+l)\psi\right]. \qquad (8)$$

Due to known relation for Bessel functions $J_\nu$ with integer index [19]

$$\int_{-\pi}^{\pi} \exp[i(h\cos\phi - \nu\phi)] d\phi = 2\pi i^{|\nu|} J_{|\nu|}(h),$$

Eq. (7) can be reduced to the form

$$u_{lm}(\rho,\psi,z) = \frac{1}{\zeta\sqrt{|m|!}} e^{i(l+m)\psi} (-i)^{|l|+|m|+1} \exp\left(\frac{i}{2\zeta}\rho^2\right) \int_0^\infty \rho_a^{|m|+1} \exp\left[-\frac{\rho_a^2}{2}\left(1-\frac{i}{\zeta}\right)\right] J_{|l+m|}\left(\frac{\rho_a\rho}{\zeta}\right) d\rho_a. \qquad (9)$$

The latter integral can be evaluated by employing the Tables of integrals (see [20], pt. 2.12.9) and the Kummer transformation [19,21] with final result

$$u_{lm}(\rho,\psi,z) = \frac{(-i)^{|l+m|+1}}{\sqrt{|m|!}} 2^{\frac{|m|-|m+l|}{2}} \frac{\Gamma\left(\frac{|m|+|m+l|}{2}+1\right)}{\Gamma(|m+l|+1)} \zeta^{\frac{|m|-|m+l|}{2}} (\zeta-i)^{-\frac{|m|+|m+l|}{2}-1}$$

$$\times e^{i(m+l)\psi} \rho^{|m+l|} \exp\left[-\frac{\rho^2}{2(1+i\zeta)}\right] M\left(\frac{|m+l|-|m|}{2}, |m+l|+1; \frac{\rho^2}{2\zeta(\zeta-i)}\right) \qquad (10)$$

where $M$ is a symbol of the confluent hypergeometric (Kummer) function [19,21]. Note that in accordance with the incident beam normalization (6) and due to the supposed 100% efficiency of the transformation, at any $m$ and $l$ the total power of the beam (10) is the same,

$$\int_0^\infty |u_{lm}(\rho,\psi,z)|^2 \rho d\rho \int_0^{2\pi} d\psi = \pi \qquad (11)$$

(this fact will be expedient for comparative study of beams with different $m$ and $l$).

The case of incident Gaussian beam considered in Ref. [12] results from (6) – (10) if $m = 0$. This equality favored to simplifying the complex amplitude expression which thus could be reduced to a combination of the Bessel functions with half-integer indices [12]. In the general case of (10) such a simplification is impossible so the beams obtained as a result of diffraction of LG modes in the CGH fully justify the name "Kummer beams" introduced earlier [9–11]. On the other hand, representation (10) is a special case of the "Hypergeometric-Gaussian" modes whose properties were recently studied in detail [22] and can be used in our further analysis.

Now consider some special and marginal cases of expression (10).

1. If $l = 0$, Eq. (10) reduces to

$$u_{0m}(\rho,\psi,z) = \frac{1}{\sqrt{|m|!}} (1+i\zeta)^{-|m|-1} e^{im\psi} \rho^{|m|} \exp\left(-\frac{\rho^2}{2(1+i\zeta)}\right), \qquad (12)$$

which quite expectedly describes the free-space evolution of the non-transformed incident beam (6).

2. Near the axis (at small $\rho$ and moderate $\zeta$ so that $(\rho/\zeta) \ll 1$) the approximate complex amplitude distribution is provided by the multiplier of (10) preceding the Kummer function. As could be expected, the field intensity appears to be proportional to $\rho^{2|m+l|}$ in agreement to the fact that the beam carries the OV of the order $|m + l|$.

3. The law of the amplitude falloff at the beam periphery $\rho \to \infty$ follows from asymptotic behavior of the Kummer function [21]

$$u_{lm}(\rho,\psi,\zeta)\big|_{\rho\to\infty} \approx \frac{(-i)^{|m+l|+1}}{\sqrt{|m|!}} \frac{\Gamma\left(\frac{|m|+|m+l|}{2}+1\right)}{\Gamma\left(\frac{|m+l|-|m|}{2}\right)} e^{i(m+l)\psi}(2\zeta)^{|m|+1}\rho^{-(|m|+2)}\exp\left(i\frac{\rho^2}{2\zeta}\right) \quad (13)$$

$$+\frac{1}{\sqrt{|m|!}}(-2i\zeta)^{|m|-|m+l|}(1+i\zeta)^{-|m+l|-1}e^{i(m+l)\psi}\rho^{|m|}\exp\left[-\frac{\rho^2}{2(1+i\zeta)}\right]. \quad (14)$$

The two summands in this representation are of important physical meaning. The second one (line (14)) can be interpreted as a result of the "regular" propagation of the initial wave (8) (compare to Eq. (12)), the first one (line (13)) describes a divergent wave with spherical wavefront component, formally outgoing from the point $\rho = \zeta = 0$, i.e. from the bifurcation point. So, just like in case of the non-vortex incident beam [12], the field of the diffracted beam generally consists of the "regular" diffracted wave, generated by the series of grooves, and of the "singular" diverging wave (in this context, word "singular" relates not to the phase singularity, which is present in both waves, but rather to the wave origin). The "singular" wave (13) is a sort of edge wave [23] appearing due to diffraction at the bifurcation point which is singular for the given CGH structure (in essence, the bifurcation point is an analog of the point singularity of a spiral phase plate equivalent to the CGH [12,24]). In particular, at $m = 0$, $l \neq 0$ Eq. (13) reduces to the asymptotic expression obtained earlier for the CGH-produced OV beams (Eq. (20) of Ref. [12]). The interference between the "regular" and "singular" waves produces the ripple structure well seen both in the amplitude and phase distributions of the near field ($\zeta \ll 1$) beam patterns presented in Figs. 2 and 3. The nature and main features of the ripple oscillations were considered elsewhere [12,24] so now we only remark that the ripples appearing in transformations of beams with already existing OV are, in general, less intensive and vanish at earlier stages of the beam evolution than was observed in the process of OV generation [12,13]. This can be attributed to the fact that, due to the factor $\rho^{-(|m|+2)}$ in Eq. (13), the diverging wave magnitude diminishes with growing $|m|$, which, in turn, is related to the vanishing incident beam amplitude at the bifurcation point.

In reality, numerical analysis is almost unable to distinguish the ripple modulations for $|m| > 1$. In Fig. 3 the structures of diffracted beams with the same resultant topological charge $m + l$ obtained from different incident OV beams are compared. It is seen that in case $(m + l) = 1 + 2$ the ripple oscillations are clearly present in the amplitude as well as in the phase distributions, while for the topologically equivalent beam with $(m + l) = 2 + 1$ the ripples are practically suppressed.

In the above asymptotic relation, at large enough $\rho$, the term (13) usually dominates provided that it does not vanish, that is unless condition

$$\frac{|m|-|m+l|}{2} = 0,1,2,... \quad (15)$$

is true. In this case, in contrast to the incident beam (6), whose amplitude falls down exponentially, the diffracted beam shows much slower power-law falloff (however, for non-zero $m$ this circumstance is not so drastic as in the case of OV generation [12] and does not prevent the existence of the second-order intensity moments [12,26,27]). An interesting detail is that the term (13) disappears (and, correspondingly, the diverging "singular" wave is not excited at all) if condition (15) does hold. This feature did not occur in the previously considered case $m = 0$ [12] so it deserves a special attention.

To disclose its physical contents, note that condition (15) is equivalent to three requirements fulfilled simultaneously: (i) $m$ and $l$ are of opposite signs, (ii) $0 \leq |l| \leq 2|m|$ and (iii) $l$ is even. Requirements (i) and (ii) mean that the absolute topological charge of the OV in the diffracted beam is not higher than that of the incident beam, i.e. the beam transformation in the CGH is accompanied by conservation (when $l = 0$ or $l = -2m$) or lowering of the absolute OV order (including its complete elimination when $l = -m$); in other words, it does not increase the "strength" of singularity. To find the meaning of requirement (iii), let us represent the function (8), describing the initial complex amplitude distribution of the diffracted beam, in the form

$$u_{lm}(\xi,\eta,0) = \frac{1}{\sqrt{|m|!}} \left(\sqrt{\xi^2 + \eta^2}\right)^{|m|} \left(\frac{\xi \pm i\eta}{\sqrt{\xi^2 + \eta^2}}\right)^{|m+l|} \exp\left(-\frac{\xi^2 + \eta^2}{2}\right) \quad (16)$$

where $\xi = \rho\cos\psi$, $\eta = \rho\sin\psi$. This function is generally singular at $\xi = \eta = 0$, where its $(|m|+1)$-th derivatives with respect to $\xi$ and $\eta$ are discontinuous. It is this discontinuity that serves a source of the "singular" diffraction wave expressed by Eq. (13). Note that the divergent wave associated with the bifurcation point emerges despite that the incident beam field in this point vanishes: discontinuities of the field derivatives can also be sources of the edge waves [23]. However, point $\xi = \eta = 0$ is not always singular. Exclusions just occur under condition (15): then function (16) is everywhere regular and infinitely differentiable. That is why in this case no edge wave emerges and the diffracted beam propagates preserving its exponential confinement. Additionally, one can easily make sure that when Eq. (15) is valid, function (8) or (16) can be represented as a superposition of Hermite-Gaussian modes, or LG modes with non-zero radial indices, that are known to obey the self-similar propagation law [18], i.e. preserve their functional form, especially, the Gaussian envelope.

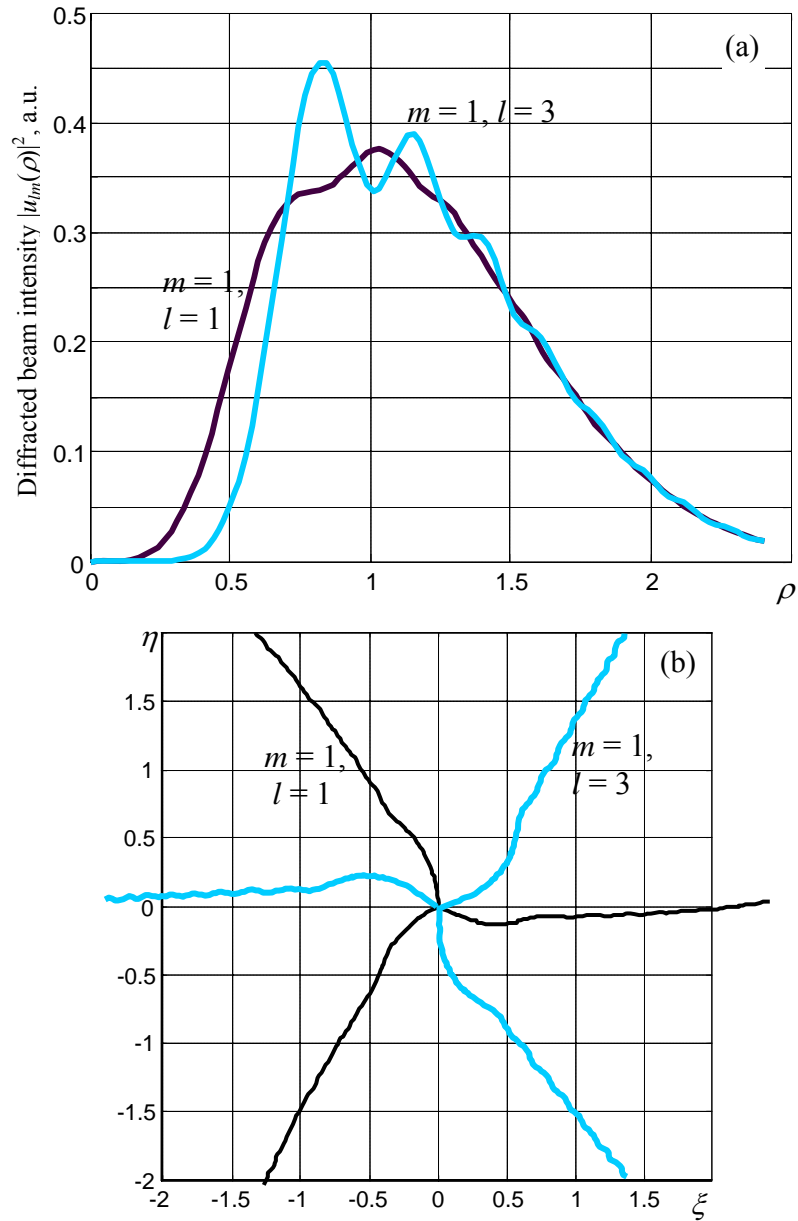

Fig. 2. (a) Radial intensity profiles and (b) equiphase lines within the beam cross section $\zeta = 0.05$ for the diffracted beams (10) with $(m, l) = (1, 1)$ (black curves) and $(m, l) = (1, 3)$ (light curves).

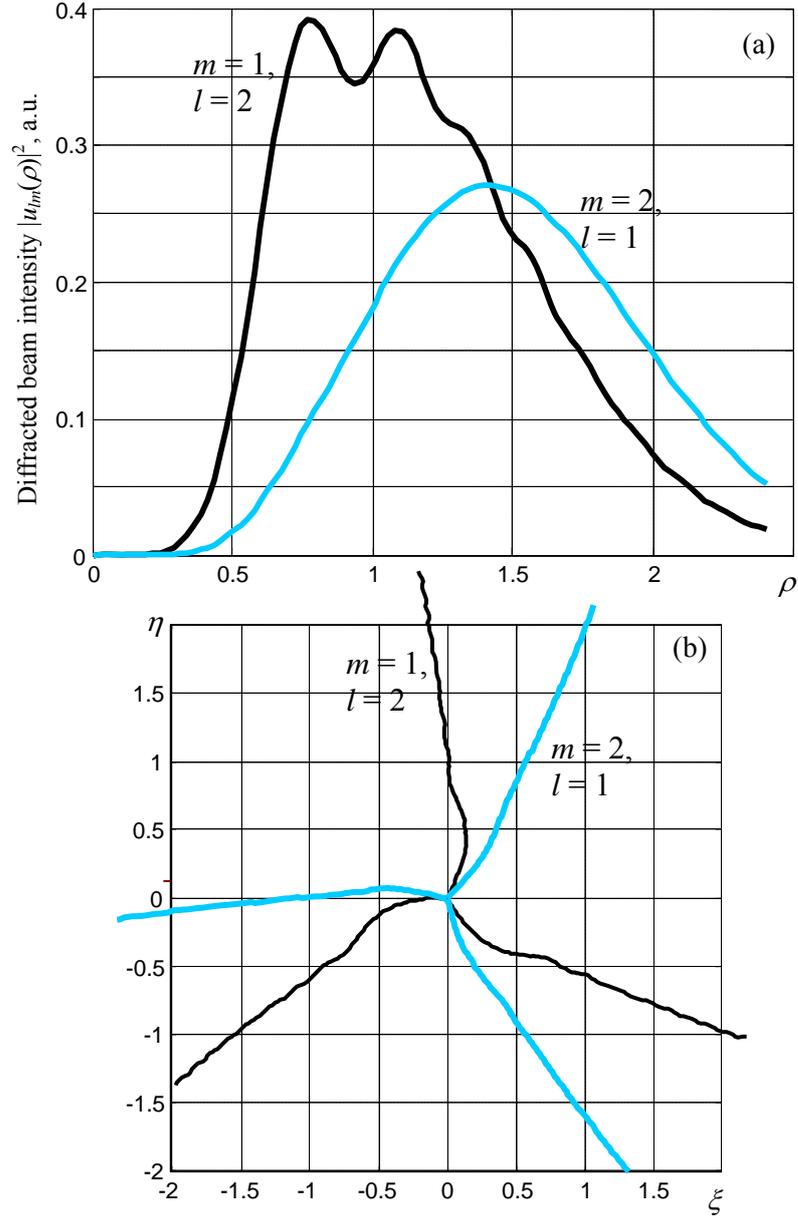

Fig. 3. (a) Radial intensity profiles and (b) equiphase lines within the beam cross section $\zeta = 0.05$ for the diffracted beams (10) with $(m, l) = (1, 2)$ (black curves) and $(m, l) = (2, 1)$ (light curves).

Now consider some special properties of the light beams created by the CGH-induced transformation of incident OV beams. The first example is associated with the problem of removing the OV from the beam. This situation occurs when an $m$-charged OV diffracts on the CGH under conditions providing $l = -m$ and is sometimes used for the OV analysis. If there exists several possible diffraction orders, the diffracted beam for which $n\mu = l = -m$ has the resulting azimuthal index $|l + m| = 0$ and no axial zero of amplitude. It is expected be an analog of the Gaussian mode and thus to possess the minimal width and divergence so it is the only diffracted beam that can be coupled to an appropriate single-mode filter [14–16].

Evolution of the beam pattern in such conditions is illustrated by Fig. 4 (remember that due to Eq. (11) all the curves in Figs. 2a, 3a, 4 and 5 describe beams with the same power). Curves in Fig. 4a show how the initial singularity existing in the incident beam at $\rho = 0$ is progressively "healed" by diffraction. It is seen that removal of the phase singularity leads to expected elimination of the axial amplitude zero. Since this process evolves with different "rates" depending on the initial singularity "strength" $|m|$, one can observe its consecutive stages. Curves 1 and 2 clearly show how the axial intensity dip is gradually filled and the axial peak is developed instead. For higher values of $|m|$ the process of "dip fillup" is just initiated: the curves for $|m| > 2$ almost repeat the radial profiles of corresponding incident beams, except the near-axis region where their values weakly differ from zero, as is shown in the inset. At moderate distances (Fig. 4b) central peaks are already well developed for all curves; however, at large $|m|$ they are not yet dominant and "compete" with off-axial maximums – a rudiments of the incident "doughnut" patterns. Rather surprisingly, maximum of the curve 1 turns out to be higher than that of curve 0: it appears that if the CGH removes the OV from the $LG_{01}$ mode, the resulting beam shows even better energy concentration than the "genuine" Gaussian beam.

This fact is not occasional and the same tendency in full measure manifests itself in the far field (Fig. 4c). Here all beams with "healed" OV demonstrate higher axial intensity compared to the Gaussian beam, and the effect grows with $|m|$. This fact can likely be applied for creation of beams with high axial energy concentration. Note that the beams corresponding to even $m$, along with high axial concentration, are well confined in the transverse direction due to the exponential amplitude decay at $\rho \to \infty$ (see the notes below Eq. (15)).

Similar features can be found in the behavior of low-order OV beams created from the high-order incident $LG_{0m}$ modes (Fig. 5). Since in this case the beam transformation is not accompanied by such impressive alteration of the beam shape as filling the intensity gap, the near-field intensity distributions (Fig. 5a) look qualitatively almost identical to the incident beam profiles. In Fig. 5b one can see the intermediate stage of the beam profile transformation. Interestingly to note, the transition from the wide "ring" of the incident high-order LG mode to the resultant narrow ring of the first-order OV beam is performed not via the ring contraction but rather via emerging a new narrow ring and subsequent energy redistribution from the wide ring to the narrow one (this is especially well seen by curves 4 – 7). At last, in the far field (Fig. 5c), almost whole energy of the beams is concentrated in the narrow rings which, again, turn out to be closer to the beam axis than the ring of the $LG_{01}$ mode – a "prototype" single-charged OV beam (curve 0). The bright ring radius decreases and the corresponding intensity maximum increases with $|m|$, that is, with growing "strength" of the beam transformation performed by the CGH.

Note that visual minimums of curves 1 – 6 in Figs. 4c and 5c never reach the horizontal axis, which is shown in the insets of the respective panels. This fact complies with the known property of the hypergeometric-Gaussian functions that can possess zero values only at $\rho = 0$ and at the infinity [22].

**Conclusion**

Now we can resume the main outcome of this paper. The general theory of the light beam transformation in CGH with embedded phase singularity [12] has been applied to the case of the incident $LG_{0m}$ mode – a standard model of a multicharged OV. Closed analytical results are obtained for the case of small-angle diffraction when approximation $\cos\theta = 1$ is valid, and the output beam preserves the circular symmetry of the incident one. The complex amplitude distribution of the diffracted beam is expressed via confluent hypergeometric (Kummer) functions so the beams generated by the CGH from $LG_{0m}$ modes belong to the enhanced the family of

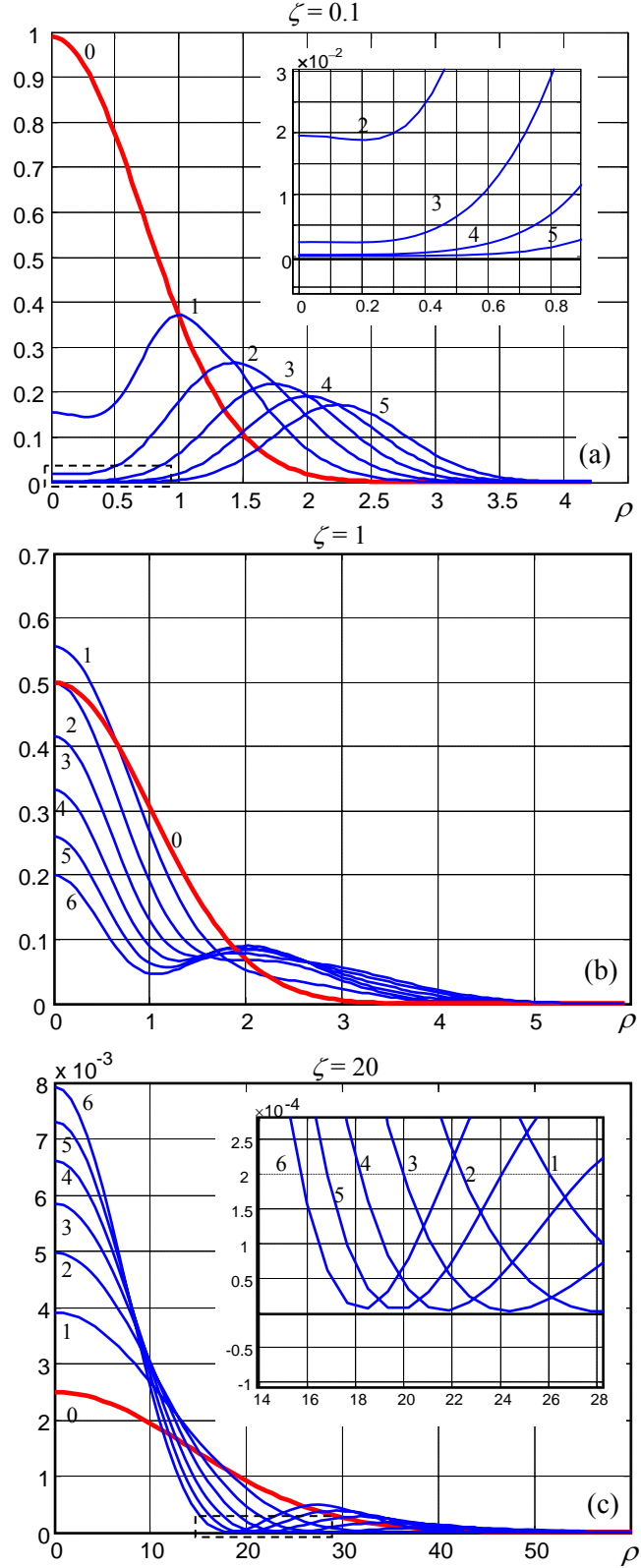

Fig. 4. Radial intensity profiles of the diffracted beams with $|l + m| = 0$ (the result of the CGH-induced transformation with complete removal of the incident beam OV and formation of the central intensity maximum): (a) in the near field ($\zeta = 0.1$); (b) in the middle field ($\zeta = 1$); (c) in the far field ($\zeta = 20$). Each curve is marked by corresponding value of $|l|$, bold curves (0) describe the current profile of the Gaussian beam (formula (6) at $m = 0$). The insets in panels (a) and (c) present magnifications of the dashed rectangle areas.

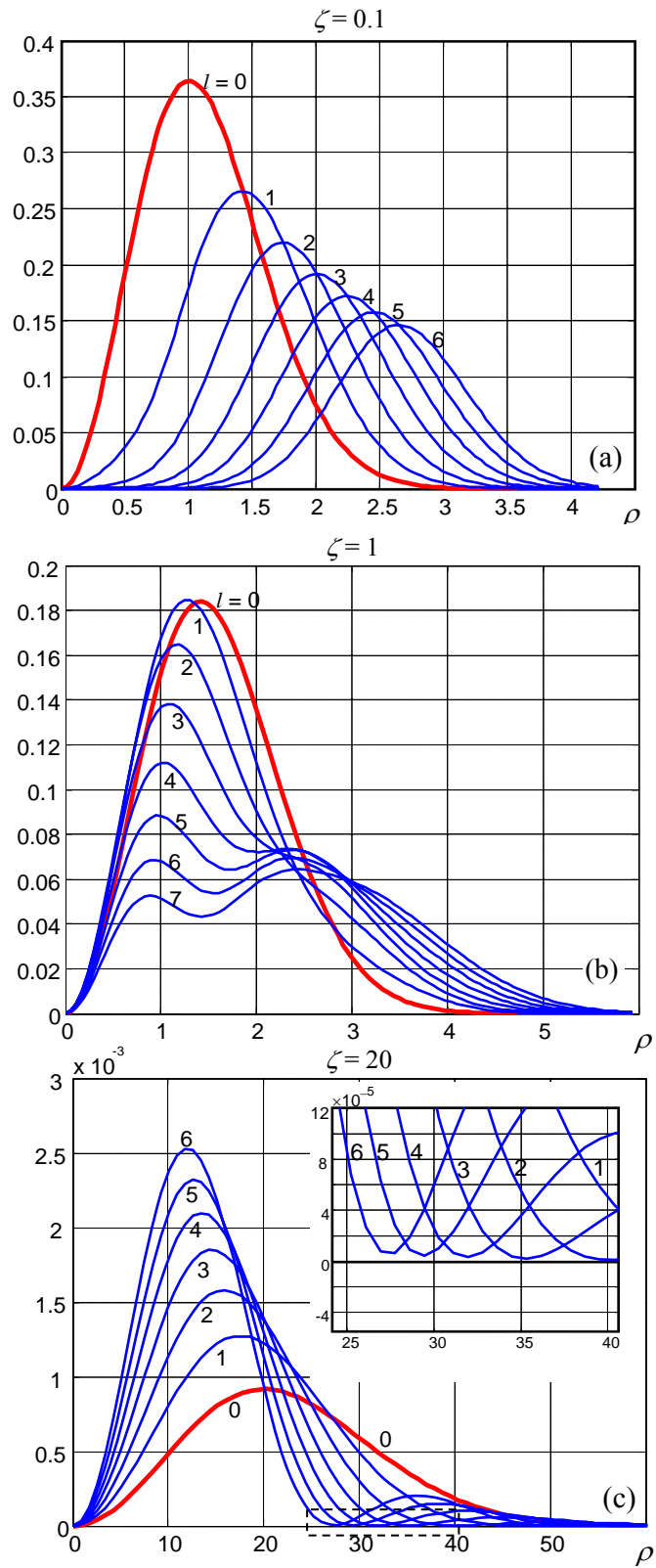

Fig. 5. Radial intensity profiles of the diffracted beams with $|l + m| = 1$, $|m| - |l| = 1$ (CGH-induced transformation with reducing the absolute OV order from $|m|$ to 1): (a) in the near field ($\zeta = 0.1$); (b) in the middle field ($\zeta = 1$); (c) in the far field ($\zeta = 20$). Each curve is marked by corresponding value of $|l|$, bold curves (0) describe the current profile of the $LG_{01}$ beam (formula (6) at $m = 1$). The insets in panels (a) and (c) present magnifications of the dashed rectangle areas.

Kummer beams [9,11–13]; on the other hand, they are special cases of the hypergeometric-Gaussian beams [22]. This identification essentially facilitates analysis of the diffracted beams' properties.

Physically, a diffracted beam is formed as a result of interference between the "regular" diffracted wave and the "singular" divergent wave that originates from the imaginary point coinciding with the hologram center (bifurcation point). This feature is common with the case of incident Gaussian beam [12,13] where the singular wave emerges due to discontinuity of the diffracted beam amplitude in the initial cross section. However, now the divergent wave is excited due to discontinuity of derivatives of the incident beam profile, rather than of the profile itself, and possesses much lower magnitude rapidly decreasing with growing topological charge of the incident beam OV $|m|$. Therefore, the interference ripples are only seen at small $|m|$ and in the near field behind the hologram, and they are less articulated and disappear at earlier stages of the beam evolution than in case $|m| = 0$.

Also, the divergent-wave component manifests itself in the fact that at the beam periphery ($\rho \to \infty$), the amplitude falls down as $\rho^{-(|m|+2)}$ instead of the exponential falloff in the incident beam. When the diffracted beam singularity is "weaker" than the incident beam singularity, at certain conditions (Eq. (15)), the divergent wave is not excited at all, and all its consequences do not exist.

If the "fork" hologram is used for "healing" the incident beam singularity (topological charge of the output beam $|l + m| < |m|$), there occur situations when the diffracted beam shows better concentration of energy near the propagation axis than the "true" $LG_{0l}$ beam. In particular, if $|l + m| = 0$, the central intensity peak of the diffracted beam can be higher and narrower than the peak of a Gaussian beam whose initial distribution coincides with the Gaussian envelope of the incident $LG_{0m}$ beam.

The results can be used for synthesis of beams with special amplitude distributions and for analyzing the spatial structure and topological charge of the OV beams. However, their realization requires additional efforts in studying the CGH action, in particular, effects of the incident beam misalignment (spatial and angular deviations of the incident beam axis from the nominal position – axis $z_a$, see Fig. 1) and consequences of the output beam symmetry breakdown caused by the diffracted beam deflection when condition $\cos\theta = 1$ is no longer valid. Corresponding generalization of the presented approach should be grounded on the substantial numerical investigations which we are planning to describe elsewhere.